\documentclass[showpacs, oneside, twocolumn, prd, amsmath, amssymb, nofootinbib]{revtex4}
\usepackage{cases}
\usepackage{amsmath}
\usepackage{amssymb}
\usepackage{amsfonts}
\usepackage{amssymb}
\usepackage{dcolumn}
\usepackage{bm}
\usepackage{bbm}
\usepackage{graphicx}
\usepackage{xcolor}
\usepackage{array}
\usepackage{subfigure}
\usepackage{hyperref}

\newcommand{\be}{\begin{equation}}
\newcommand{\ee}{\end{equation}}
\newcommand{\ba}{\begin{eqnarray}}
\newcommand{\ea}{\end{eqnarray}}

\newcommand{\gsim}{\mathrel{\hbox{\rlap{\lower.55ex \hbox {$\sim$}}
                   \kern-.3em \raise.4ex \hbox{$>$}}}}
\newcommand{\lsim}{\mathrel{\hbox{\rlap{\lower.55ex \hbox {$\sim$}}
                   \kern-.3em \raise.4ex \hbox{$<$}}}}

\def\roughly#1{\mathrel{\raise.3ex\hbox{$#1$\kern-.75em%
\lower1ex\hbox{$\sim$}}}}
\def\lsim{\roughly<}
\def\gsim{\roughly>}

\def\({\left(}
\def\){\right)}
\def\[{\left[}
\def\]{\right]}
\def\<{\langle}
\def\>{\rangle}

\newcommand{\omits}[1]{}

\hypersetup{colorlinks=true,
            breaklinks=true,
            pdfstartview=Fit,
            linkcolor=blue,
            citecolor=blue,
            urlcolor=blue}

\usepackage{color}

\begin{document}

\title{Note on the Green's function formalism and topological invariants}

\author{
Yehao Zhou$^{1,2}$\footnote{yzhou3@perimeterinstitute.ca}
Junyu Liu$^{3}$\footnote{jliu2@caltech.edu}
}

%\author{Yi-Fu Cai$^{2}$} \footnote{yifucai@ustc.edu.cn}
%\author{Shu Lin$^{1}$} \footnote{linshu8@mail.sysu.edu.cn}
%\author{Junyu Liu$^{2, 3}$} \footnote{jliu2@caltech.edu}
%\author{Jia-Rui Sun$^{1}$} \footnote{sunjiarui@sysu.edu.cn}

\affiliation{1) Perimeter Institute for Theoretical Physics, Waterloo, ON N2L 2Y5, Canada}
\affiliation{2) Department of Physics \& Astronomy, University of Waterloo, Waterloo, ON N2L 3G1, Canada}
\affiliation{3) Walter Burke Institute for Theoretical Physics, California Institute of Technology, Pasadena, CA 91125, United States}

\begin{abstract}
It has been discovered previously that the topological order parameter could be identified from the topological data of the Green's function, namely the (generalized) TKNN invariant in general dimensions, for both non-interacting and interacting systems. In this note, we show that this phenomenon has a clear geometric derivation. This proposal could be regarded as an alternative proof for the identification of the corresponding topological invariant and the topological order parameter.
\end{abstract}

\maketitle

\section{Introduction}
Recent researches in condensed matter community show novel applications of quantum field theory method in topological materials, which are not typically distinguished by local ordered parameters, but by topological quantum numbers, which could be described and computed using topological field theories in the low energy limit (for review of those works, for instance, see \cite{Hasan:2010xy,Qi:2011zya}, and also a standard book \cite{BH}).
\\
\\
Among various applications of quantum field theory, one notable example is the existence of topological insulators (see \cite{Hasan:2010xy,Qi:2011zya,BH,Kane:2004bvs,Kane:2005zz,Fu:2007uya,Moore:2006pjk}). The topological insulator is a special kind of quantum material, where the corresponding bulk band gap is similar to an ordinary insulator, but the boundary (edge or surface) of the material has been equipped with a state which is protected by the topological numbers of the bulk geometry. As a simple example, the (2+1)-dimensional noninteracting topological insulator is described by the Thouless-Kohmoto-Nightingale-den Nijs (TKNN) invariant \cite{Thouless:1982zz}, defined by the first Chern class of complex bundle over the Brillouin zone. In the Chern-Simons theory description, one can identify the TKNN invariant with the level of the Chern-Simons action, which is proportional to the Hall conductivity. This identification could be proven by exact computations (for instance, see \cite{Thouless:1982zz}) or geometric arguments (for instance, see \cite{Witten:2015aoa}).
\\
\\
On the other hand, given the action of the Chern-Simons effective theory, one could derive an inversion formula from the field contents to the level (Hall conductivity) of the Chern-Simons action. This inversion formula is typically evaluated from a corresponding one-loop Feynman diagram \cite{Niemi:1983rq,Golterman:1992ub,Qi:2008ew}, as an integration of a combination of Green's functions over all spacetime dimension, where we call it as the topological order parameter. Thus, based on the knowledge of both the identification between the TKNN invariant and the level of the Chern-Simons theory, and the identification between the level and the topological order parameter, we arrive at the conclusion that the TKNN invariant has been identified with the topological order parameter, which could also be shown by straightforward calculation. The topological order parameter is very useful and has strong computational power when generalized to the interacting case, which only requires the Green's function data instead of a complex bundle over Hilbert space for the TKNN invariant. In the past research \cite{OR1,OR2,OR3}, people show that the identification between the topological order parameter and a generalized version of the TKNN invariant still holds in the interacting theory, by using smooth homotopy from the interacting Green's function to a non-interacting theory (see also, several related discussions \cite{Wang:2010xh,rela,Ryu:2012he,Martin-Ruiz:2015skg}).
\\
\\
In this note, we will describe a geometric proof for identification between the (generalized) TKNN invariant and the topological order parameter without explicit computation. This identification does not need the Hall conductivity (the level of the Chern-Simons action) in the middle, and it could form a clear geometric picture while those two quantities are the same based on knowledge of algebraic topology.
\\
\\
The organization of this note is given as following. In Section \ref{back} we will review some preparations towards our mean result. In Section \ref{geo} we will provide the geometric proof. In Section \ref{nogo} we will provide a no-go theorem and an extension based on the identification in the previous discussions. In Section \ref{conc} we will arrive at a conclusion and outlook. In Appendix \ref{A}, we will review some basic knowledge of the Chern-Simons forms for completeness.
\section{Preparation}\label{back}
\subsection{The Green's function on a momentum manifold}
In this section we will write down the setup of the problem. We consider an even ($2n$) spatial dimensional material, with the momentum space defined as the manifold $M$ with the coordinate $k$. The functions over momentum space are often reduced to Brillouin zone because of periodicity of the crystal, which should be understood as quotients of momentum space following the argument from the Bloch theorem. Thus as standard knowledge, the overall manifold $M$ is often regarded as torus. However, here our statement could cover the generic, compact manifold $M$.
\\
\\
To understand the physical property of electronic dynamics in a material, we often define the Matsubara Green's function as (thermal) two point correlations. Here, we are working in zero temperature, mathematically speaking, the Matsubara Green's function $G(i\omega,k)$ is a mapping from momentum spacetime $M\times \mathbb{R}$ to the group $\text{GL}(N,\mathbb{C})$, where $N$ is the number of bands. Here we assume it to be finite (the notation  $\omega$ we use here for frequency is often called $i\omega$ in other literatures),
\begin{align}
G(\omega ,k):M\times \mathbb{R}\mapsto \text{GL}(N,\mathbb{C})~.
\end{align}
In the non-interacting theory, the Green's function is written as
\begin{align}
G(\omega ,k)=\frac{1}{\omega -H(k)}~,
\end{align}
where $H(k)$ is the Hamiltonian, the mapping from manifold $M$ to Hermitian matrices in $\text{GL}(N,\mathbb{C})$. In the electronic system, the Green's function is written in the Kallen-Lehmann form
\begin{align}
&{{G}_{\alpha \beta }}(\omega ,k)=\nonumber\\
&\sum\limits_{m}{\left[ \frac{\left\langle  0 \right|{{c}_{k\alpha }}\left| m \right\rangle \left\langle  m \right|c_{k\beta }^{\dagger }\left| 0 \right\rangle }{\omega -({{E}_{m}}-{{E}_{0}})}+\frac{\left\langle  m \right|{{c}_{k\alpha }}\left| 0 \right\rangle \left\langle  0 \right|c_{k\beta }^{\dagger }\left| m \right\rangle }{\omega +({{E}_{m}}-{{E}_{0}})} \right]}~,
\end{align}
where $m$ labels the eigenvector of $H$ (or generically, $H-\mu N_p$ with chemical potential $\mu$ and particle number $N_p$) with vacuum $0$, and $c,c^\dagger$s are annihalation and creation operators of fermions, labeled by momentum $k$ and band $\alpha$.
\\
\\
Ideally, there might be infinitely large number of bands and we should take an infinite sum. However, practically people will take a cutoff and choose a periodic boundary condition. In this case, we have a finite abelian group, while finite choices of $\alpha$ mean generators of it. That is why we wish to say that $N$ is the number of bands (there is only a finite number of Brillouin zones).
\\
\\
For future convenience, we will obtain the following decomposition formula for the Green's function. We can decompose the Green's function as
\begin{align}
G=\frac{G+{{G}^{\dagger }}}{2}+i\frac{G-{{G}^{\dagger }}}{2i}~,
\end{align}
where two terms, without the factor $i$, are both Hermitian. Using Kallen-Lehmann form we have
\begin{align}
  & G(\omega ,k)=\nonumber\\
  &\sum\limits_{m}{\left[ \frac{\text{Re}(\omega )-({{E}_{m}}-{{E}_{0}})}{|\omega -({{E}_{m}}-{{E}_{0}}){{|}^{2}}}u_{m}^{\dagger }{{u}_{m}}+\frac{\text{Re}(\omega )+({{E}_{m}}-{{E}_{0}})}{|\omega +({{E}_{m}}-{{E}_{0}}){{|}^{2}}}v_{m}^{\dagger }{{v}_{m}} \right]} \nonumber\\
 & -i\sum\limits_{m}{\left[ \frac{\text{Im}(\omega )}{|\omega -({{E}_{m}}-{{E}_{0}}){{|}^{2}}}u_{m}^{\dagger }{{u}_{m}}+\frac{\text{Im}(\omega )}{|\omega +({{E}_{m}}-{{E}_{0}}){{|}^{2}}}v_{m}^{\dagger }{{v}_{m}} \right]}~.
\end{align}
Here ${{u}_{m}}=({{u}_{m,\alpha }})=\left\langle  m \right|c_{k\alpha }^{\dagger }\left| 0 \right\rangle $ and ${{v}_{m}}=({{v}_{m,\alpha }})=\left\langle  0 \right|c_{k\alpha }^{\dagger }\left| m \right\rangle $ are vectors labelled by $\alpha$. Collecting terms we have
\begin{align}
  & G=\bar{\omega }\sum\limits_{m}{\left[ \frac{u_{m}^{\dagger }{{u}_{m}}}{|\omega -({{E}_{m}}-{{E}_{0}}){{|}^{2}}}+\frac{v_{m}^{\dagger }{{v}_{m}}}{|\omega +({{E}_{m}}-{{E}_{0}}){{|}^{2}}} \right]}+ \nonumber\\
 & \sum\limits_{m}{\left[ \frac{({{E}_{m}}-{{E}_{0}})v_{m}^{\dagger }{{v}_{m}}}{|\omega +({{E}_{m}}-{{E}_{0}}){{|}^{2}}}-\frac{({{E}_{m}}-{{E}_{0}})u_{m}^{\dagger }{{u}_{m}}}{|\omega -({{E}_{m}}-{{E}_{0}}){{|}^{2}}} \right]}~.
\end{align}
Thus, this formula shows that we could decompose the Green's function as
\begin{align}
G(\omega ,k)=\bar{\omega }A(\omega ,k)+B(\omega ,k)~,
\end{align}
where
\begin{align}
  & A(\omega ,k)\equiv \frac{u_{m}^{\dagger }{{u}_{m}}}{|\omega -({{E}_{m}}-{{E}_{0}}){{|}^{2}}}+\frac{v_{m}^{\dagger }{{v}_{m}}}{|\omega +({{E}_{m}}-{{E}_{0}}){{|}^{2}}} \nonumber\\
 & B(\omega ,k)\equiv \frac{({{E}_{m}}-{{E}_{0}})v_{m}^{\dagger }{{v}_{m}}}{|\omega +({{E}_{m}}-{{E}_{0}}){{|}^{2}}}-\frac{({{E}_{m}}-{{E}_{0}})u_{m}^{\dagger }{{u}_{m}}}{|\omega -({{E}_{m}}-{{E}_{0}}){{|}^{2}}}~,
\end{align}
and we observe the fact that $u_mu_m^\dagger$ and $v_mv_m^\dagger$ are Hermitian and positive semidefinite. So we get the desired decomposition of $G$ with the property that $A(\omega ,k)$ and $B(\omega ,k)$ are Hermitian and $A(\omega ,k)$ is positive definite. In fact, we claim that $A$ is positive definite rather than being only positive semidefinite. The argument is that the Green's function is non-degenerate for any $(\omega,k)\in \mathbb R\times M$, suppose that $A$ is degenerate at some $(\omega,k)$ for some vector $r$, namely, $A(\omega,k)\cdot r=0$, then it's obvious that $r$ is orthogonal to every $u_m$ and $v_m$, by definition of $A$, hence $B(\omega,k)\cdot r=0$, by definition of $B$, which contradicts the fact that $G$ is non-degenerate.
\subsection{The topological order parameter}
In this part we will define topological numbers we will use. Firstly, on a $2n$ dimensional compact manifold $M$, with a Matsubara Green's function $G$, the topological order parameter is defined by
\begin{align}\label{TOP}
{{\mathcal{N}}_{2n}}={{\left( \frac{1}{2\pi i} \right)}^{n+1}}\frac{n!}{(2n+1)!}\int_{M\times \mathbb{R}}{{{\mathcal{G}}^{*}}}\text{Tr}({{\eta }^{2n+1}})~,
\end{align}
where $\eta$ is the fundamental one form on the Lie group $\text{GL}(N,\mathbb C)$\footnote{In fact, $\text{Tr}(\eta^{2n+1} )$ can be identified with the generator of the rational homotopy group $\pi _{2n+1}(\text{GL}(N,\mathbb C))_{\mathbb Q}=\mathbb Q$ when $N$ is large enough (more precisely $N>n$), or equivalently the generator $x_{2n+1}$ in the cohomology $\text{H}^*(\text{GL}(N,\mathbb C),\mathbb Q)=\Lambda ^*[x_1,x_3,\cdots,x_{2N-1}]$.}, namely, $\eta_g=g^{-1}dg$ and $\mathcal{G}$ is the inverse of the Matsubara Green's function.
\\
\\
Now we are going to show that the integral in the above definition is convergent. To proceed, we need the following observation for the behaviour of the Green's function when $\omega$ approaching infinity. Following the previous expansion formula, and as we claimed before, there is only a finite sum in the expansion so that we can extract $1/|\omega|^2$ from the fraction and control the rest by $\mathcal O\left( \omega^{-2}\right)$, we have
\begin{align}
G(\omega ,k)= \frac{1}{\omega }{{A}_{0}}(k)+\mathcal{O}\left( \frac{1}{{{\omega }^{2}}} \right)~,
\end{align}
where
\begin{align}
A_{0}(k)= \sum_{m}u^{\dagger}_m u_m+v^{\dagger}_m v_m~,
\end{align}
is positive definite, and then apply the differential operator we have
\begin{align}
&dG(\omega ,k)=\frac{1}{\omega }d{{A}_{0}}(k)+\mathcal{O}\left( \frac{1}{{{\omega }^{2}}} \right)dk\nonumber\\
&-\frac{{{A}_{0}}(k)}{{{\omega }^{2}}}d\omega +\mathcal{O}\left( \frac{1}{{{\omega }^{3}}} \right)d\omega~,
\end{align}
where $dk$ is the short hand notation for differential forms in $M$.
\\
\\
Now we could claim that the integral
\begin{align}
{{\left( \frac{1}{2\pi i} \right)}^{n+1}}\frac{n!}{(2n+1)!}\int_{M\times \mathbb{R}}{{{\mathcal{G}}^{*}}}\text{Tr}({{\eta }^{2n+1}})~,
\end{align}
is convergent, where $\mathcal{G}$ is the inverse of $G$, and the integral is taken by integrating on $M$ slice first. To show this, it's not harmful to use $G$ rather than $\mathcal{G}$ in the integrand, which only changes the sign (because $g^{-1}dg=-d(g^{-1})g$). Now in the large $\omega$ limit, from the result above, we have an asymptotic expansion
\begin{align}
{{G}^{-1}}dG=A_{0}^{-1}d{{A}_{0}}+\mathcal{O}\left( \frac{1}{\omega } \right)dk-\frac{d\omega }{\omega }+\mathcal{O}\left( \frac{1}{{{\omega }^{2}}} \right)d\omega~.
\end{align}
Since terms involving $d\omega/\omega^s$ for $s>1$ are automatically convergent, we only need to look at $d\omega/\omega$ terms, which are proportional to
\begin{align}
\int_{|\omega |\ge C}{\frac{d\omega }{\omega }}\int_{M}{\text{Tr}}[{{(A_{0}^{-1}d{{A}_{0}})}^{2n}}]~.
\end{align}
But $\text{Tr}[(A_{0}^{-1}dA_{0})^{2n} ]$ vanishes because $\text{Tr}(AB)=\text{Tr}(BA)$ and we can move the first $A_{0}^{-1}dA_{0}$ term to the tail and get a minus sign
\begin{align}
\text{Tr}[{{(A_{0}^{-1}d{{A}_{0}})}^{2n}}]=-\text{Tr}[{{(A_{0}^{-1}d{{A}_{0}})}^{2n}}]~.
\end{align}
Thus the result is convergent.
\subsection{Passing to contour integral}
\begin{figure}
\centering
\includegraphics[width=0.3\textwidth]{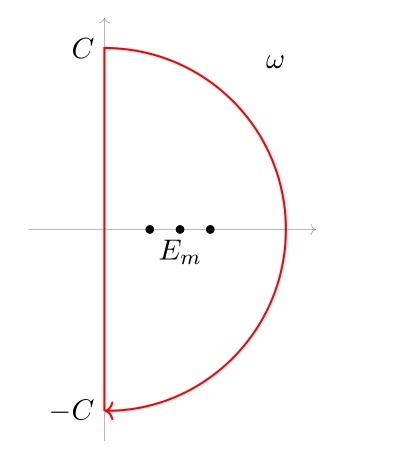}
\caption{The integration contour for $\omega$ in the complex plane $\mathbb{C}$.} \label{fig:Contour}
\end{figure}

As is mentioned above, Eq.\ref{TOP} looks like a topological invariant, but it's defined on a non-compact manifold, and it is not compactly supported, so we need to find a substitute to characterize it in a more familiar way. The substitution is given in the following. We pick a cutoff by integrating $\omega$ in a segment $[-C,C]$ ($C$ is sufficiently large to avoid hitting $E_m$), and then close the segments by connecting $C$ with $-C$ via a semicircle in the complex plane, as is shown in the Figure \ref{fig:Contour}.
A possible issue is that, is the analytical continued Green's function still non-degenerate along this semicircle or not? This is guaranteed by the large $\omega$ expansion of $G$
\begin{align}
G(\omega ,k)=\frac{1}{\omega }{{A}_{0}}(k)+\mathcal{O}\left( \frac{1}{{{\omega }^{2}}} \right)~.
\end{align}
Although we only show the expansion in the case that $\omega$ is restricted in the imaginary line, there is no difficulty to extend the argument to the case that $\omega$ runs through a large circle. So for large enough $C$, the analytical continued Green's function is non-degenerate along the semicircle.
\\
\\
We will also claim that the integral along the semicircle will vanish as $C\to \infty$. This can be proved by looking at the asymptotic behaviour of the integrand at infinity. Note that for extension of $dG$, there will be some $d\bar {\omega}$ terms, but they are suppressed by $1/|\omega|^2$. Then we can apply the same arguments from the convergence statement to conclude that the integral along the big circle will approach zero as $C$ goes to infinity.
\\
\\
Thus,  we can safely reduce the original integral which involves $\pm i \infty$ to an integral on the compact manifold $M\times T$ ($T$ for torus). An important consequence is that, this is a well-defined topological invariant, so it is a constant as we move the circle to the infinity, which converges to the original integral, thus we can effectively use a contour to replace the imaginary line.
\subsection{Smooth homotopy to the non-interacting theory}
Since the integral of the pull-back cocycle on the compact manifold $M\times T$ is stable under the homotopy of maps $G_u:M\times T\times [0,1]\to \text{GL}(N,\mathbb C)$ with $u\in [0,1]$, we can deform the original Green's function to be more manageable. In fact, it is shown in \cite{OR1,OR2} that we can always deform it to a Green's function associated to a non-interacting theory. We shall explain this below.
\\
\\
A non-interacting theory has the standard Green's function
\begin{align}
\frac{1}{\omega -H(k)}~,
\end{align}
where $H(k)$ is the Hamiltonian. Now we can define a system with an effective Hamiltonian
\begin{align}
H(k)\equiv G^{-1}(0,k)~,
\end{align}
so that the associated Green's function is
\begin{align}
\widetilde G(\omega,k)=\frac{1}{\omega+G^{-1}(0,k)}~.
\end{align}
Here the inverse is well-defined because $\omega$ is either on the imaginary line or $|\omega|=C$ being large so that $\omega+G^{-1}(0,k)$ has nonzero eigenvalue. Connect $G$ with $\widetilde G$ via a smooth homotopy
\begin{align}
G_u(\omega,k)=(1-u) G(\omega,k)+u\widetilde G(\omega,k)~.
\end{align}
We claim that $\forall u\in [0,1]$, $G_u(\omega,k)\in \text{GL}(N,\mathbb C)$. This is obvious when $\omega=0$. When $\omega$ has a nonzero image part, we could take any vector $r$ with unit norm, and compute
\begin{align}
  & \text{Im}(\left\langle  r \right|{{G}_{u}}(\omega ,k)\left| r \right\rangle )\nonumber\\
  &=\text{Im}\left( (1-u)\left\langle  r \right|G(\omega ,k)\left| r \right\rangle  \right)\nonumber\\
  &+\text{Im}\left( u\left\langle  r \right|\frac{1}{\omega +{{G}^{-1}}(0,k)}\left| r \right\rangle  \right) \nonumber\\
 & =\text{Im}(\bar{\omega })\times\nonumber\\
 &\left( (1-u)\left\langle  r \right|A(\omega ,k)\left| r \right\rangle +\frac{u}{|\omega {{|}^{2}}+|\left\langle  r \right|{{G}^{-1}}(0,k)\left| r \right\rangle {{|}^{2}}} \right)~,
\end{align}
since $A$ is positive definite. For $\omega=C\in \mathbb R$, this comes from an asymptotic expansion
\begin{align}
{{G}_{u}}(C,k)=\frac{1-u}{C}{{A}_{0}}(k)+\frac{u}{C}\text{Id}+\mathcal{O}\left( \frac{1}{{{C}^{2}}} \right)~,
\end{align}
where $\text{Id}$ is the identity. So we conclude that
\begin{align}\label{Formula}
{\mathcal{N}_{2n}}={{\left( \frac{1}{2\pi i} \right)}^{n+1}}\frac{n!}{(2n+1)!}\int_{M\times T}{{{{\widetilde{\mathcal{G}}}}^{*}}\text{Tr}({{\eta }^{2n+1}})}~,
\end{align}
where $\widetilde {\mathcal{G}}(\omega,k)=\omega +G^{-1}(0,k)$.
\subsection{The generalized TKNN invariant}
Generically, the generalized TKNN invariant will include the topological description for interacting system \cite{OR1,OR2,OR3}. Here we define it and make a brief introduction.
\\
\\
Firstly, we notice that the Green's function with zero frequency $G(0,k)$ is Hermitian and non-degenerate, so we can split the total space in to a direct sum of two parts
\begin{align}
E(k)={{E}_{+}}(k)\oplus {{E}_{-}}(k)~,
\end{align}
where $E(k)$ is the total space and $E_\pm (k)$ corresponds to the subspace spanned by corresponding plus or minus eigenvalues. As we change the momentum $k\in M$, those $E_\pm (k)$ patch together to form two complex vector bundles on $M$. We call them $E_\pm$. There is a corresponding splitting of bundles
\begin{align}\label{split}
E=E_+\oplus E_-~.
\end{align}
Note that $E$ is a trivial bundle (a function on $M$ with vector values). In the original papers \cite{OR1,OR2}, they are called R-space ($E_+$) and L-space ($E_-$) because they are formed by zeroes in the right hand side and left hand side of the origin.
\\
\\
Now we define the generalized TKNN invariant $C_n$ to be the $n\text{'th}$ Chern character of $E_+$
\begin{align}\label{split}
C_n:=\langle \text{ch}_n(E_+),[M]\rangle~.
\end{align}
\section{Main proof}\label{geo}
In this section, we will provide a geometric proof of the theorem
\begin{align}\label{main}
\mathcal{N}_{2n}=C_n~,
\end{align}
as a simple equation between the generalized TKNN invariant and the topological order parameter. The proof will be organized in the following steps. Firstly, we can interpretate the formula Eq.\ref{Formula} in terms of an integration over a Chern-Simons form on $M\times T$, and then a characteristic class of a complex vector bundle on $M\times T\times T$. Secondly, we should go to another perspective that we view this bundle as a direct sum of two bundles, where each of them comes from a tensor product of a bundle on $M$ with a line bundle on $ T\times T$. Finally, we will do some calculations using the property of Chern characters.
\\
\\
The first step is basically contained in the Appendix \ref{A}. According to example in this appendix, the topological order parameter $\mathcal{N}_{2n}$ equals to the minus of the $(n+1)\text{'th}$ Chern character of the twisted bundle on $M\times T\times T$, which is constructed by gluing the trivial bundle $\underline {\mathbb C}^N$ on $M\times T\times \{0\}$ and $M\times T\times \{1\}$ via $\widetilde{\mathcal G}(\omega, k)=\omega +G(0,k)$. We will denote this bundle $\widetilde{E}$, so we claim that the topological order parameter is given by the following simple characteristic class
\begin{align}
\mathcal{N}_{2n}=-\langle \text{ch}_{n+1}(\widetilde{E}),[M\times T\times T]\rangle~.
\end{align}
We can also continuously deform the gluing map $\widetilde{\mathcal G}(\omega, k)$ and obtain an isomorphic twisted bundle because of following isomorphisms:
\begin{align}
  & \left\{ P\in \text{Bu}{{\text{n}}_{\text{U(N)}}}(M\times T\times T)|P\text{ trivial on }M\times T\times \left\{ 0 \right\} \right\} \nonumber\\
 & \simeq \left[ (M\times T\times T,M\times T\times \left\{ 0 \right\}),(\text{B}U(N),*) \right] \nonumber\\
 & \simeq \left[ S(M\times T),\text{B}U(N) \right] \nonumber\\
 & \simeq \left[ M\times T,U(N) \right]~.
\end{align}
The first isomorphism is the definition of classifying space, the second comes from the fact that $\pi_1(\text{B}U(N))$ is trivial so that we can contract $M\times T\times \left\{ 0 \right\}\vee T$ and obtain the suspension of $M\times T$, and the third isomorphism is just $\Omega \text{B}G\simeq G$.
\subsection{Continuous deformation of $\tilde{\mathcal{G}}$}
Now we know that the diagonalization of the Green's function gives a spitting
\begin{align}
\widetilde{\mathcal G}\in \Gamma\left(M\times T,\text{GL}(E'_+)\oplus  \text{GL}(E'_-)\right)~,
\end{align}
where $E'_{\pm}$ is the pull-back of $E_{\pm}$ on $M$. We can parametrize the torus $T$ by $\theta\in[-1,1]$, and we claim that
\begin{align}
\widetilde{\mathcal G} \simeq \text{Id}_{E'_+}\oplus e^{-i\pi \theta}\text{Id}_{E'_-}~,
\end{align}
and this homotopy is taken in $\Gamma\left(M\times T,\text{GL}(E'_+)\oplus  \text{GL}(E'_-)\right)$.
\\
\\
In fact, we firstly notice that if $\omega$ lies on the semicircle contour, then $\omega+A$ and $\omega+B$ are homotopic if $A$ and $B$ are both positive definite or negative definite: consider the path $\omega+(1-\lambda)A+\lambda B$ and any vector $\left\langle  r \right|$, $\left\langle  r \right|\omega +(1-\lambda )A+\lambda B\left| r \right\rangle \ne 0$ whenever $\omega$ is pure imaginary, and if $\omega$ has large modulus, this is also nonzero because of the domination of $\omega$, so $\omega+(1-\lambda)A+\lambda B$ is non-degenerate for any $\lambda \in [0,1]$.
\\
\\
Applying this observation to $\widetilde{\mathcal G}|_{E'_+}$ and $\widetilde{\mathcal G}|_{E'_-}$ separately, we conclude that
\begin{align}
\widetilde{\mathcal G}(\omega,k) \simeq (\omega+1)\text{Id}_{E'_+}\oplus(\omega-1)\text{Id}_{E'_-}~,
\end{align}
where $\omega$ runs through the semicircle contour described in Figure \ref{fig:Contour}. Then we can shrink the semicircle associated to $\omega+1$ to a constant function $1$, and shrink the semicircle associated to $\omega-1$ to the unit circle $e^{-i\pi \theta},\theta\in[-1,1]$ (we cannot shrink the contour to a constant like the $E'_+$ sector because the contour can not pass through the singularity at zero), as is shown in the following Figure \ref{fig:def}. Hence the claim is proved.
\begin{figure}
\centering
\includegraphics[width=0.3\textwidth]{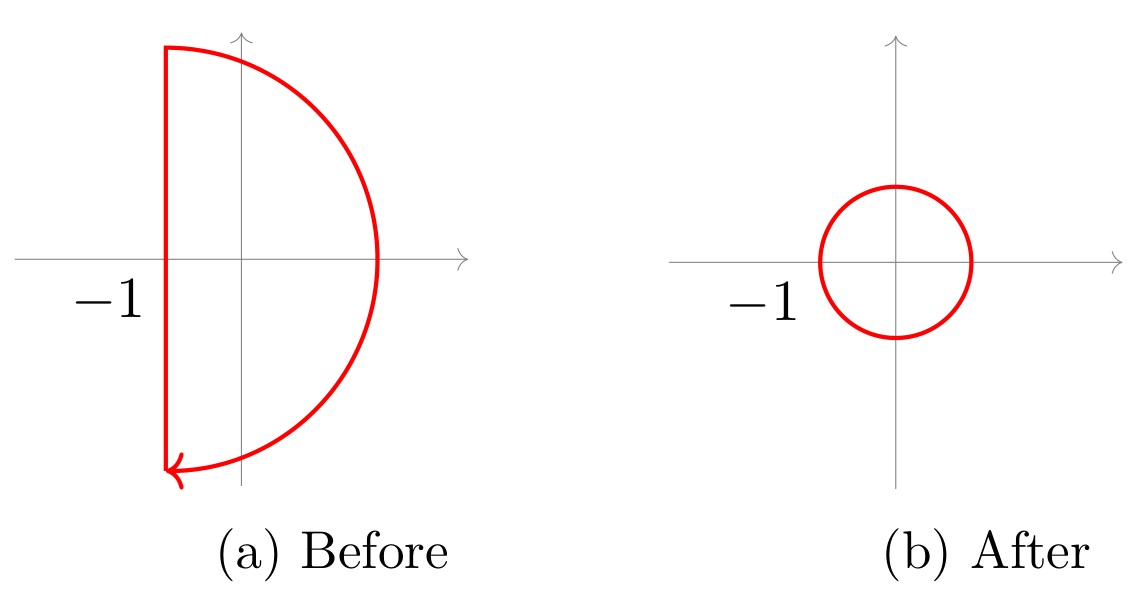}
\caption{Deformation for $E'_-$ sector.} \label{fig:def}
\end{figure}
\subsection{Another perspective for $\tilde{E}$}
Now $\widetilde E$ is constructed by gluing the trivial bundle $\underline {\mathbb C}^N$ on $M\times T\times \left\{ 0 \right\}$ and $M\times T\times \left\{ 1 \right\}$ via
\begin{align}
\text{Id}_{E'_+}\oplus e^{-i\pi \theta}\text{Id}_{E'_-}~.
\end{align}
There is a simple observation: the twisting data, namely, the gluing function is only relevant to the coordinate of $T$, which is a clue that we can \emph{split} the base manifold $M$ from the construction. As a matter of fact, we could claim that, the twisted bundle $\widetilde E$ can be equivalently constructed by taking the trivial bundle $\underline{\mathbb C}^N=E_+\oplus E_-$ on $M$, and twisting the plus and minus components by tensoring with two different line bundles on 2-torus $T^2$, namely
\begin{align}
\widetilde E\simeq E_+\boxtimes \underline {\mathbb C}\oplus E_-\boxtimes \mathcal L~,
\end{align}
where $\underline {\mathbb C}$ is the trivial line bundle, and $\mathcal L$ is defined by gluing the boundary of trivial line bundle on $T\times [0,1]$ via $e^{-i\pi \theta}$. Here we use the box product to denote the externel tensor product of bundles, more precisely, if manifolds $M$ and $N$ have bundles $F$ and $G$ respectively, then $F\boxtimes G$ is the bundle $\pi_M^*F\otimes \pi_N^*G$ on $M\times N$, via a pull-back along projection $\pi_{M,N}$ to corresponding components.
\\
\\
In fact, since $\text{Id}_{E'_+}\oplus e^{-i\pi \theta}\text{Id}_{E'_-}$ leaves the two components $E'_+$ and $E'_-$ invariant, we have a splitting
\begin{align}
\widetilde E=\widetilde E_+\oplus \widetilde E_-~,
\end{align}
where $\widetilde E_+$ is the gluing of the bundle $E'_+$ on $M\times T$ with a $\underline {\mathbb C}$ on $T$, but $E'_+$ is the tensor product of $E_+$ with a $\underline {\mathbb C}$ on $T$, so
\begin{align}
\widetilde E_+=E_+\boxtimes \underline {\mathbb C}~.
\end{align}
On the other hand, $\widetilde E_-$ is the twist of $E'_-$ by $e^{-i\pi \theta}$ at the point $(\theta,k)$, or equivalently, forming the line bundle $\mathcal L$ on $T^2$ first, and tensor product with $E_-$, which gives
\begin{align}
\widetilde E_-= E_-\boxtimes \mathcal L~,
\end{align}
thus we prove the claim.
\subsection{Conclude the proof}
To compute $\mathcal{N}_{2n}$, it suffices to calculate the Chern character of $\widetilde E$, according to the previous results, we have
\begin{align}\label{Chern}
\text{ch}(\widetilde E)&=\text{ch}(E_+\boxtimes \underline {\mathbb C})+\text{ch}(E_-\boxtimes \mathcal L)\nonumber\\
&=\text{ch}(E_+)\cup  \text{ch}(\underline {\mathbb C})+\text{ch}(E_-)\cup \text{ch}(\mathcal L)\nonumber\\
&=p^*\text{ch}(E_+)+\text{ch}(E_-)\cup \text{ch}(\mathcal L)~.
\end{align}
Here $p$ is the projection from $M\times T^2$ to $M$. After contracting with the fundamental cycle $[M\times T^2]$, according to K\"{u}nneth theorem, only the $\langle\text{ch}_n(E_-)\cup \text{ch}_1(\mathcal L),[M\times T^2]\rangle$ survives, which in turn equals to
\begin{align}
\langle\text{ch}_n(E_-) ,[M]\rangle\langle\text{ch}_1(\mathcal L),[ T^2]\rangle~.
\end{align}
Moreover, $E_+\oplus E_-$ is a trivial bundle on $M$, so
\begin{align}
\mathcal{N}_{2n}&=-\langle\text{ch}_n(E_-) ,[M]\rangle\langle\text{ch}_1(\mathcal L),[ T^2]\rangle\nonumber\\
&=\langle\text{ch}_n(E_+) ,[M]\rangle\langle\text{ch}_1(\mathcal L),[ T^2]\rangle\nonumber\\
&=\langle\text{ch}_1(\mathcal L),[ T^2]\rangle C_n~.
\end{align}
Thus we need to show that
\begin{align}\label{test}
\langle\text{ch}_1(\mathcal L),[ T^2]\rangle=1~.
\end{align}
In fact, recall the construction of $\mathcal L$, it is the gluing of the boundary of a trivial line bundle on $T\times [-1,1]_{\theta}$ via $e^{-i\pi \theta}$. Use the exmaple in the Appendix \ref{A} again, we conclude that
\begin{align}
\langle\text{ch}_1(\mathcal L),[ T^2]\rangle&=-\left(\frac{1}{2\pi i}\right)\int_{-1}^1 e^{i\pi \theta}de^{-i\pi \theta}\nonumber\\
&=\frac{1}{2}\int_{-1}^1d\theta\nonumber\\
&=1~,
\end{align}
which concludes the proof.
\section{No-go theorem and extension}\label{nogo}
\begin{figure}
\centering
\includegraphics[width=0.2\textwidth]{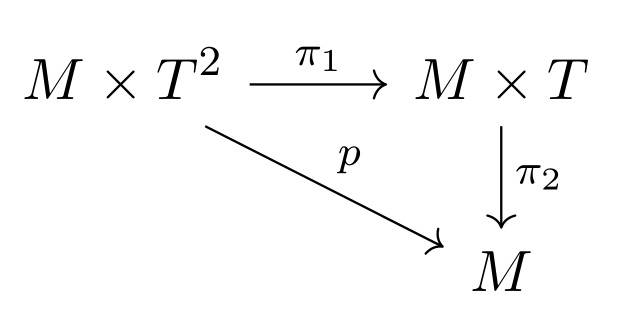}
\caption{A commutative diagram.} \label{fig:com}
\end{figure}

Based on the technology we have built above, we could discuss some extensions of the identification theorem. An obvious attempt to extend this topological order parameter is looking for similiar construction such as
\begin{align}
\int_{M\times \mathbb R}\mathcal G^*\left(\text{Tr}(\eta^{n_1}) \text{Tr}(\eta^{n_2})\cdots \text{Tr}(\eta^{n_l})\right)\text{ , } n_i\text{ are odd}~,
\end{align}
or more formally, the pull-back via $\mathcal G$ of other elements in $\text{H}^{n_1+n_2\cdots +n_l}(\text{GL}(N,\mathbb C))$. Here $n_i$ are odd because $\text{Tr}(\eta^{2k})$ vanishes:
\begin{align}
&\text{Tr}(\eta_{a_1}\eta_{a_2}\cdots\eta_{a_{2k}})dx^{a_1}\wedge dx^{a_2}\cdots \wedge dx^{a_{2k}}\nonumber\\
&=\text{Tr}(\eta_{a_2k}\eta_{a_1}\cdots\eta_{a_{2k-1}})dx^{a_1}\wedge dx^{a_2}\cdots \wedge dx^{a_{2k}}\nonumber\\
&=-\text{Tr}(\eta_{a_{2k}}\eta_{a_1}\cdots\eta_{a_{2k-1}})dx^{a_{2k}}\wedge dx^{a_1}\cdots \wedge dx^{a_{2k-1}}\nonumber\\
&=-\text{Tr}(\eta_{a_1}\eta_{a_2}\cdots\eta_{a_{2k}})dx^{a_1}\wedge dx^{a_2}\cdots \wedge dx^{a_{2k}}~.
\end{align} 
It can be shown in a similar way with the method in this note that they are convergent. However, it turns out that they are trivial. Before we show that, we shall define a refined version of theorem \ref{main} first, which relates the Chern character of $E_+$ (or equivalently $E_+$) to the pull-back  of $\text{H}^*(\text{GL}(N,\mathbb C))$.
\\
\\
Consider the following commutative diagram \ref{fig:com}.
Use formula Eq.\ref{Chern} and also the result Eq.\ref{test} we obtain
\begin{align}
p_!(\text{ch}(\widetilde E))&=p_!(p^*\text{ch}(E_+)+\text{ch}(E_-)\cup \text{ch}(\mathcal L))\nonumber\\
&=\text{ch}(E_-)~.
\end{align}
On the other hand, the commutative diagram and the nature of Chern-Simons forms gives
\begin{align}
p_!(\text{ch}(\widetilde E))&={\pi_2}_!\circ {\pi_1}_!(\text{ch}(\widetilde E))\nonumber\\
&=-{\pi_2}_!\left(\mathcal G^*\sum_{i\ge 0} x_{2i+1}\right)~.
\end{align}
Here the lower shriek symbol means integrating along fibers, and $x_{2i+1}$ is the generator of $\text{H}^*(\text{GL}(N,\mathbb C))$ defined by $\text{Tr}(\eta^{2i+1})$ (up to some constant, which is not important here). Thus we have proved the following refined version of the main theorem: The Chern character of $E_+$ is related to $\mathcal G^*\text{H}^*(\text{GL}(N,\mathbb C))$ by
\begin{align}
\text{ch}(E_+)={\pi_2}_!\left(\mathcal G^*\sum_{i\ge 0} x_{2i+1}\right)~,
\end{align}
or more transparently,
\begin{align}
\text{ch}_{l}(E_+)=\left(\frac{1}{2\pi i}\right)^l\frac{l!}{(2l+1)!}\int _{-i\infty}^{i\infty}d\omega\text{ } \mathcal G^*\text{Tr}(\eta^{2l+1})~.
\end{align}
This equation holds in the cohomology $\text{H}^{2l}(M)$.
\\
\\
Another conclusion from the diagram Figure \ref{fig:com} is that
\begin{align}
{\mathcal G}^* \sum_{i\ge 0} x_{2i+1} &= -{\pi_1}_!(\text{ch}(\widetilde E))\nonumber\\
&=-{\pi_1}_!({\pi_1}^*{\pi_2}^*\text{ch}(E_-)\cup \text{ch}(\mathcal L))\nonumber\\
&=-{\pi_2}^*\text{ch}(E_-)\cup {\pi_1}_!\text{ch}(\mathcal L)~.
\end{align}
We have already known from the conclusion Eq.\ref{test} that $\text{ch}(\mathcal L)=1+\alpha_1\cup \alpha_2$ where $\alpha_{1,2}$ is the generator of $\text{H}^1(T,\mathbb Z)$ for the first and second 1-torus respectively, so this gives us the inverse map of the above theorem
\begin{align}
\left(\frac{1}{2\pi i}\right)^l\frac{l!}{(2l+1)!}\text{ } \widetilde{\mathcal G}^*\text{Tr}(\eta^{2l+1})=-\alpha_2\cup {\pi_2}^*\text{ch}_l(E_-)~.
\end{align}
This equation holds in the cohomology $\text{H}^{2l+1}(M\times T)$. Then the promised no-go theorem is that, the naive generalization to the topological order parameter
\begin{align}
&\int_{M\times i \mathbb R}\mathcal G^*\left(\text{Tr}(\eta^{n_1}) \text{Tr}(\eta^{n_2})\cdots \text{Tr}(\eta^{n_l})\right)\nonumber\\
&\text{ , } n_1+n_2+\cdots+n_l=2n+1~,
\end{align}
for $l>1$ are trivial, following from the fact that every cup product of two \emph{trace-blocks}
\begin{align}
\mathcal G^*\text{Tr}(\eta^{n_1})\cup \mathcal G^*\text{Tr}(\eta^{n_2})~,
\end{align}
is trivial, because there are two $\alpha_2$ involved, and they square to zero. Physically, it means that the above simple manipulations could not give any new topological information of the electronic system that are beyond the identification theorem in the main proof. 
\\
\\
To see a nontrivial generalization, instead of taking wedge product (or cup product in cohomology) first and doing integration along the imaginary line, we can integrate out $\omega$ first, and take the wedge product in $M$. Then we could claim that, there exist generalizations to the topological order parameter
\begin{align}
&N'_{j_1,j_2\cdots j_l}:=\left(\frac{1}{2\pi i}\right)^n \left(\prod_l\frac{{j_l}!}{(2j_l+1)!}\right)\times\nonumber\\
&\int_M \left(\prod_l\int_{-i\infty}^{i\infty}d\omega_l\text{ }\mathcal G(\omega_l,n)^*\text{Tr}(\eta^{2j_l+1})\right)~,
\end{align}
which is related to the topology of the band structure by
\begin{align}
N'_{j_1,j_2\cdots j_l}=(-1)^l\langle \text{ch}_{j_1}(E_-)\text{ch}_{j_2}(E_-)\cdots \text{ch}_{j_l}(E_-),[M]\rangle~.
\end{align}
This extension has experimental implications: the extension $N'_{j_1,j_2\cdots j_l}$, understood as Chern characters, could be also interpreted as extensions of TKNN invariants. On the other hand, it has a closed form expression by the Green's functions and thus it could be measured experimentally. 

\section{Conclusion and outlook}\label{conc}
In this paper, we mainly discuss the identification proof of the topological order parameter and the (generalized) TKNN invariant motivated by topological condensed matter physics. The main part of this proof is a geometric identification of the non-interacting theory from the topological order parameter to the TKNN invariant defined over filled bands. This proof is an alternative way to find an inversion formula of the Chern-Simons level using classical field data from Feynman diagram, combining the argument that the TKNN invariant from band theory could be identified as the level of the Chern-Simons theory. Through this proof, we build up some intuitions from statements of algebraic topology about what happens between those two quantities and what happens to form such a connection. Although we are motivated by condensed matter physics, this is a statement which is intrinsically from the low energy Chern-Simons theory. Thus, based on such a construction, we revisit and formalize the smooth contour deformation that is used in \cite{OR1,OR2} by an order estimation of the Green's function using spectral decomposition. The deformation will make the identification theorem more general, from non-interacting theory to interacting theory. Moreover, we discuss some possible extensions over this identification, to make the picture more fruitful based on the technologies we use.
\\
\\
We believe that the geometric interpretation of those identities might give a clear and conceptual connection between two different topological invariants, in general even dimensional compact space at once, instead of brute force computations. Moreover, about the smooth homotopy, the new proof does not use the physical statement given by \cite{OR2}. Instead, by rigorous mathematical treatments restricting the homotopy performed in a compact space, the proof ensures a legitimate contour integral. Finally, our geometric interpretation brings a novel no-go theorem and a natural extension from the original formula. 
\\
\\
As an outlook, we will comment on following possible working directions inspired by our result
\begin{itemize}
\item It would be interesting to generalize those discussions to the Fractional Quantum Hall Effects.
\item We observe that this proof is technically pretty similar with proofs in historical discussions in the structure of gauge and gravitational anomalies \cite{AlvarezGaume:1984dr}. It is possible to find more physical consequences of this work and possible extensions from high energy theory to condensed matter physics.
\item Recently, there are rising interests about the mathematical structures related to the Green's functions on the Hilbert space of band Hamiltonians. For instance \cite{A1,A2}. \cite{A1} proves localization theorems of the Green's function: in a Chern insulator the Green's function cannot decay too fast if the Hamiltonian is finite-range. Moreover, the Green's function is predicted not to be finite-range if the quantum Hall response is nontrivial. \cite{A2} proves a no-go statement that for certain types of Hamiltonians, the Integer or Fractional Quantum Hall Effects are always trivial. It might be interesting to see if our technology could shed light on those researches and prove more novel generic and interesting claims.
\end{itemize}

\subsection*{Acknowledgments}
A significant part of this work is contributed to the partial fulfilment of \emph{Perimeter Scholars International} program of YZ. This work is initiated from a discussion between Jingxiang Wu and YZ. We thank Xie Chen, Kevin Costello and Davide Gaiotto for helpful communications. Research at Perimeter Institute is supported by the Government of Canada through Industry Canada and by the Province of Ontario through the Ministry of Research and Innovation. JL acknowledges support from the U.S. Department of Energy, Office of Science, Office of High Energy Physics, under Award Number DE-SC0011632.

\appendix
\section{On the Chern-Simons forms}\label{A}
We briefly review the Chern-Weil homomorphism and Chern-Simons forms based on the original paper \cite{Chern:1974ft} for completeness, and write a crucial example that could be used in the main text.
\\
\\
For a manifold $X$ with a principal $G$-bundle $P$ on $X$, and a connection $\omega$ on $P$ with curvature $F$, we can use $G$-invariant polynomials on the Lie algebra $g$ to construct characteristic differential forms. Namely, pick an $S\in \text{Sym}^r(g^*)^G$, we can view $S(F)$ as a differential $2r$-form via sending an antisymmetrized tensor product of $2r$ vectors to $g\otimes \cdots \otimes g$ first, and then acted on by $S$. One can show that $S(F)\in \Omega ^{2r}(P)$ is closed and invariant under local gauge transformation, so it can descend to a closed $2r$-form on the base manifold $X$. More importantly, the cohomology class associated to $S(F)$ (on $X$) is independent of the connection. This is called the Chern-Weil construction.
\\
\\
Suppose that there are two connections $\omega _0$ and $\omega_1$ on $P$ with curvature $F_0$ and $F_1$, then the cohomology class of $S(F_1)-S(F_0)$ is trivial, and by Hodge theory there exists a $2r-1$-form with exterior differential equals to $S(F_1)-S(F_0)$. It was shown by Chern and Simons that this $(2r-1)$-form can be constructed in the following canonical way.
\\
\\
We can take the product space $X\times [0,1]$ and extend the bundle trivially, put the connection $\omega_0$ and $\omega_1$ on two slices $X\times \left\{0\right\}$ and $X\times \left\{1\right\}$ respectively, and extend them via a smooth curve $\omega_t$, for example $\omega_t=(1-t)\omega_0+t\omega_1$, so the curvature on $P\times \left[0,1 \right]$ is
\begin{align}
F=F_t+\frac{\partial \omega_t}{\partial t}dt~,
\end{align}
where $F_t$ is the curvature on each slice. From basic calculus
\begin{align}
S(F_1)-S(F_0)=\int_0^1 dt\mathcal L_{\partial t}S(F)~,
\end{align}
so we find the desired $(2r-1)$-form which is called the Chern-Simons form
\begin{align}
\text{CS}_{2r-1}(S;\omega_0,\omega_1):=\int_0^1 dt \text{ }i_{\partial t}S(F)~.
\end{align}
it can be easily shown that the Chern-Simons form is gauge invariant thus descending to the base $X$, still satisfy the defining property
\begin{align}
S(F_1)-S(F_0)=d\text{CS}_{2r-1}(S;\omega_0,\omega_1)~.
\end{align}
Actually, Chern-Simons form is independent of the smooth curve $\omega_t$, up to a coboundary term. This can be easily observed from the fact that the Chern-Weil homomorphism is independent of the connection, up to a coboundary term, so a second curve $\omega'_t$ gives another $S(F')$ on $X\times [0,1]$ but the difference between them is $d\eta$, where $\eta\in \Omega^{2r-1}(P)^G$ (or equivalently, $\Omega^{2r-1}(X)$). The difference between Chern-Simons forms is
\begin{align}
\int_0^1 dt \text{ }i_{\partial t}d\eta=\eta_1-\eta_0-d\left(\int_0^1 dt \text{ }i_{\partial t}\eta\right)~,
\end{align}
but $\eta_1=\eta_0=0$ because the starting and ending points are fixed.
\\
\\
So without losing information at the cohomology level, we can choose a special curve
\begin{align}
\omega_t=(1-t)\omega_0+t\omega_1~,
\end{align}
to obtain a more explicit formula
\begin{align}
r\cdot \int_0^1 dt \text{ }S\left(\omega_1-\omega_0,F_t,\cdots,F_t\right)~,
\end{align}
and from the construction, we also have this additivity property for Chern-Simons forms
\begin{align}
&\text{CS}_{2r-1}(S;\omega_0,\omega_2)\nonumber\\
&=\text{CS}_{2r-1}(S;\omega_0,\omega_1)+\text{CS}_{2r-1}(S;\omega_1,\omega_2)~.
\end{align}
If the base manifold $X$ has odd dimension $2n-1$, $\text{CS}_{2r-1}(S;\omega_0,\omega_1)$ is a top form and can be integrated on $X$ and gives
\begin{align}
\int_X\text{CS}_{2r-1}(S;\omega_0,\omega_1)=\int_{X\times [0,1]}S(F)~.
\end{align}
Moreover, If $\omega_1$ is a gauge transformation of $\omega_0$, namely, $\omega_0^h$ for some $G$-valued function $h$, so that we can glue the bundle at $X\times\{0\}$ with $X\times\{1\}$ via the gauge transformation $h$, the above integral formula has a simple topological interpretation: the integration of
$\text{CS}_{2r-1}(S;\omega_0,\omega_1)$ on $X$ gives a characteristic number of the twisted bundle (mapping torus $\mathcal M_P$) on $X\times T$. This will be the key to the proof of the main theorem in this paper.
\\
\\
If the bundle $P$ is trivial, we can take the trivial connection as $\omega_0$, and any other connection $A$ (as an element in $\Omega^1(X,g)$) will generate a special Chern-Simons form
\begin{align}
\text{CS}_{2r-1}(S;A):=\text{CS}_{2r-1}(S;0,A)~.
\end{align}
Now we introduce,
\\
\\
\textbf{Example:} Take $G=\text{U}(N)$, and a trivial bundle $P$. Consider the invariant polynomials
\begin{align}
S_s(g)=\text{Tr}(g^s)/s!~.
\end{align}
By standard symmetric function argument, these polynomials generate the algebra
\begin{align}
\text{Sym}^*(\text{gl}(N)^*)^{\text{U}(N)}~,
\end{align}
then the associated characteristic classes are Chern characters\footnote{
Here we use the same $i/2\pi$ convention with topologist.}
\begin{align}
\text{ch}_s(A)=\left(\frac{i}{2\pi}\right)^s\frac{\text{Tr}(F^s)}{s!}~,
\end{align}
where $F$ is the curvature form,
\begin{align}
F=dA+A\wedge A~.
\end{align}
Then the associated Chern-Simons form
\begin{align}
\text{CS}_{2s-1}(A):=\text{CS}_{2s-1}(S_s;0,A)~,
\end{align}
is given by
\begin{align}
&\text{C}{{\text{S}}_{2s-1}}(A)=\left(\frac{i}{2\pi}\right)^s\frac{1}{(s-1)!}\times\nonumber\\
&\int_{0}^{1}{\text{Tr}\left( A\wedge {{(tdA+{{t}^{2}}A\wedge A)}^{s-1}} \right)}dt~.
\end{align}
For instance, for $s=2$ we obtain the ordinary Chern-Simons action for gauge field $A$,
\begin{align}
&\left(\frac{i}{2\pi}\right)^2\int_{0}^{1}{\text{Tr}\left( A\wedge (tdA+{{t}^{2}}A\wedge A) \right)}dt\nonumber\\
&=\frac{-1}{4\pi^2}{\text{Tr}\left( \frac{1}{2}A\wedge dA+\frac{1}{3}A\wedge A\wedge A \right)}~.
\end{align}
If $A$ is a pure gauge, namely, $A=g^{-1}dg$ for some $G$-valued function on $X$, then $F=dA+A\wedge A=0$ (Maurer-Cartan), and the Chern-Simons form can be simplified as
\begin{align}
\text{C}{{\text{S}}_{2s-1}}(A)&=\left(\frac{i}{2\pi}\right)^n\frac{\text{Tr}\left( A^{2s-1} \right)}{(s-1)!}\int_{0}^{1}(t^2-t)^{s-1}dt\nonumber\\
&=-\left(\frac{1}{2\pi i}\right)^s\frac{(s-1)!}{(2s-1)!}\text{Tr}\left( A^{2s-1} \right)~.
\end{align}
If the dimension of the manifold $X$ is $2s-1$ ($n=s$), the integration of this Chern-Simons form on $X$, following the mapping torus argument above, is equal to the $s\text{'th}$ Chern character of the mapping torus $\mathcal M_P$ associated to the gauge transformation $g$ on the boundary.

\end{document}